\documentclass[preprint,12pt]{elsarticle}

\usepackage{amssymb}

\usepackage{amsmath}
\usepackage{amsthm}
\usepackage{bm}

\newtheorem{ass}{Assumption}
\newdefinition{rmk}{Remark}
\newproof{pf}{Proof}
\newproof{pot}{Proof of Theorem \ref{thm2}}

\usepackage{graphicx}
\usepackage{enumitem}

\usepackage[table]{xcolor}

\usepackage{tabularx}
\usepackage{booktabs}
\usepackage{multirow}
\usepackage{array}
\newcolumntype{M}[1]{>{\centering\arraybackslash}m{#1}}

\usepackage{algorithm}
\usepackage{algorithmic}

\usepackage{makecell}

\journal{Engineering Applications of Artificial Intelligence}

\begin{document}

\begin{frontmatter}



\title{A new approach for combined model class selection and parameters learning for auto-regressive neural models}

\author[polimi,rse]{Corrado Sgadari}
\author[polimi]{Alessio La Bella\corref{cor1}}
\ead{alessio.labella@polimi.it}
\author[polimi]{Marcello Farina}

\affiliation[polimi]{organization={Dipartimento di Elettronica, Informazione e Bioingegneria, Politecnico di Milano},
            addressline={Piazza Leonardo da Vinci 32}, 
            city={Milano},
            postcode={20133}, 
            country={Italy}}

\affiliation[rse]{organization={Ricerca sul Sistema Energetico – RSE S.p.A.},
            addressline={Via Rubattino 54}, 
            city={Milano},
            postcode={20134}, 
            country={Italy}}

\cortext[cor1]{Corresponding author}

\begin{abstract}
This work introduces a novel approach for the joint selection of model structure and parameter learning for nonlinear dynamical systems identification. Focusing on a specific Recurrent Neural Networks (RNNs) family, i.e., Nonlinear Auto-Regressive with eXogenous inputs Echo State Networks (NARXESNs), the method allows to simultaneously select the optimal model class and learn model parameters from data through a new set-membership (SM) based procedure. The results show the effectiveness of the approach in identifying parsimonious yet accurate models suitable for control applications. Moreover, the proposed framework enables a robust training strategy that explicitly accounts for bounded measurement noise and enhances model robustness by allowing data-consistent evaluation of simulation performance during parameter learning, a process generally NP-hard for models with autoregressive components.
\end{abstract}




\end{frontmatter}

\section{Introduction}
Mathematical models are essential in engineering and science for representing the behavior of dynamical systems. Physical models are often highly detailed and complex, typically involving many parameters and equations that are often uncertain or not fully known. Such models are therefore less suitable for control-oriented applications. Among the possible modeling strategies, parametric black-box identification offers an effective alternative by assuming a predefined model structure that encapsulates system dynamics and estimates its unknown parameters \cite{bittanti_model_2019},\cite{billings_nonlinear_2013}. Unlike physical models, black-box approaches rely directly on input–output data to provide compact and accurate representations of system behavior. In this context, choosing an appropriate model structure for parametric models is crucial yet challenging. The model must be simple, intuitive, and accurately describe the system's dynamics. Poor structure selection can lead to inadequate models, whereas redundant models are prone to overfitting and significant estimation errors.
For linear models, methods such as cross-validation (CV), final prediction error (FPE) \cite{akaike_fitting_1969}, Akaike information criterion (AIC) \cite{akaike_new_1974} and minimum description length (MDL) \cite{rissanen_modeling_1978} help balance complexity and accuracy. Subspace-based techniques, such as those proposed in \cite{van2012subspace}, also provide effective tools for model order selection. However, these criteria are designed for linear systems less effective for nonlinear systems, where model accuracy and size do not straightforward correlate \cite{palumbo_seismic_2001}.

In case of nonlinear systems, universal approximators such as polynomial models \cite{jones_recursive_1989} and neural networks \cite{narendra_identification_1990} are often used for their flexibility in modeling complex nonlinearities \cite{aguirre_dynamical_1995}. Effective heuristc model selection approaches such as the Forward-Regression Orthogonal Estimator (FROE) method \cite{korenberg_orthogonal_1988}, based on the Error Reduction Ratio (ERR) criterion, allow one to iteratively select meaningful features while maintaining model parsimony. Evolutions of this method, such as the Simulation Error Minimization with Pruning (SEMP) algorithm \cite{piroddi_identification_2003}, further refine model structure selection.
More recently, automatic architecture optimization techniques for recurrent neural networks have been explored, often relying on exhaustive or probabilistic metaheuristic searches over possible configurations \cite{laddach_automatic_2022},\cite{hutter_automated_2019}. While these approaches can effectively determine performing model structures, they usually involve high computational effort and do not take into account the uncertainty affecting the data.

To address the above-mentioned issues, we propose a data-consistent heuristic procedure for the identification of nonlinear dynamical systems, with particular focus on black-box parametric nonlinear models. The method jointly performs model structure selection and parameter estimation, and explicitly accounts for bounded disturbances in the data, being grounded in set-membership identification principles. The effectiveness of this procedure is demonstrated through tests on Nonlinear Auto-Regressive with eXogenous inputs Echo State Networks (NARXESNs) \cite{damico_data-based_2023}.
The latter are flexible tools for nonlinear system identification and time series forecasting, combining the autoregressive (ARX) components of traditional models \cite{ljung_system_1999} with the dynamic memory and computational efficiency of Echo State Networks (ESNs) \cite{jaeger_echo_2001}. This hybrid architecture effectively captures temporal dependencies while maintaining low computational complexity.

The paper is structured as follows. First, Section~\ref{sec:problem_statement} formalizes the problem and the candidate model family. Then, Section~\ref{sec:iterativeModelClassSelection} presents the iterative model class selection algorithm. Section~\ref{sec:NARXESN_SM_setid} describes a new set-membership based parameter optimization method and introduces a novel performance index for model class comparison. Section~\ref{sec:SimulationResults} reports simulation results, whereas Section~\ref{sec:RealWorldCaseStudy} shows the application to a real-world case study. Finally, Section~\ref{sec:Conclusions} concludes and outlines future work.

\section{Problem Statement}\label{sec:problem_statement}

\subsection{Data Generating System}

Model class selection within the identification framework includes choosing the most suitable model class for a given system. Consider a nonlinear dynamical system $\mathcal{S}$ that belongs to the family of NARXESN models evolving as
\begin{subequations}\label{eq:NARXESN_model}
\renewcommand{\theequation}{\theparentequation.\alph{equation}}
\begin{align}
    \chi^\circ(k+1) &= \zeta^\circ\big(W_\chi^\circ\chi^\circ(k) + W_{\bm \varphi}^\circ{\bm \varphi}^\circ(k) + W_z^\circ z(k+1)\big), \label{eq:NARXESN_reservoir} \\
    \mathcal{S}:\quad z(k+1) &= W_{out_1}^\circ\chi(k) + W_{out_2}^\circ {\bm \varphi}^\circ(k), \label{eq:NARXESN_output} \\
    y(k) &= z(k) + w(k).\label{eq:NARXESN_measurement}
\end{align}
\end{subequations}
In \eqref{eq:NARXESN_model} \( \chi^\circ \in \mathbb{R}^{\nu^\circ} \) denote the state of the system, \( z \in \mathbb{R} \) is the nominal output and \( u \in \mathbb{R} \) is the system input, whereas \( y \in \mathbb{R} \) represents the measured output, which is corrupted by an additive measurement noise \( w \in \mathbb{R} \). For conciseness, the approach presented in this paper is illustrated for single–input single–output (SISO) systems, although the extension to multi–input multi–output (MIMO) systems is straightforward.
The regressor vector \( {\bm \varphi}^\circ = [\varphi_1^\circ,\, \ldots ,\,\varphi_{n_{\bm \varphi}^\circ}^\circ]^\top\in \mathbb{R}^{n^o_{\bm \varphi}} \) contains $n_{\bm \varphi}^\circ$ different elements belonging to the set $\Phi(n_z^\circ,n_u^\circ)$ defined as
\[
\Phi(n_z^\circ,n_u^\circ) := \big\{z(k) , \ldots , z(k - n^o_z + 1) , u(k), \ldots , u(k - n^o_u + 1) \big\},
\]
i.e.,  collection of past values of the nominal output and system input delayed up to $n_z^\circ$ and $n_z^\circ$, respectively. The matrices $W_\chi^\circ \in \mathbb{R}^{\nu^\circ \times \nu^\circ}$, $W_{\bm \varphi}^\circ \in \mathbb{R}^{\nu^\circ \times n_{\bm \varphi}^\circ}$, $W_z^\circ \in \mathbb{R}^{\nu^\circ}$, $W_{out_1}^\circ \in \mathbb{R}^{\nu^\circ}$, and $W_{out_2}^\circ \in \mathbb{R}^{n_{\bm \varphi}^\circ}$ are unknown, and $\zeta^\circ(\cdot): \mathbb{R}^{\nu^\circ} \rightarrow \mathbb{R}^{\nu^\circ}$ with $\zeta^\circ(\cdot)=[\zeta_1(\cdot),\ldots,\zeta_{n_{nl}^\circ}(\cdot),\ldots,\zeta_{\nu^\circ}]^\top$ is a vector of Lipschitz continuous functions applied element-wise, among which $n_{nl}^\circ$ are nonlinear (e.g., $\zeta_i(\cdot)=\tanh(\cdot)$ for $i=1,\ldots,n_{nl}^\circ$).

Note that, defining $\theta^\circ = [W_{out_1}^{\circ^\top}\; W_{out_2}^{\circ^\top}]^\top$ and $\psi^\circ(k) = [\,\chi^\circ(k)^\top\; {\bm \varphi}^\circ(k)^\top]^\top$, equation \eqref{eq:NARXESN_output} can be expressed in the linear-in-the-parameter (LITP) form:
\begin{equation}\label{eq:NARXESN_output_LITP}
z(k+1) = \theta^{\circ^\top}\psi^\circ(k).
\end{equation}

For system $\mathcal{S}$ we assume that Assumption \ref{ass:NARXESN_SM_assumptions} is verified.

\begin{ass}\label{ass:NARXESN_SM_assumptions}
    For all $k\geq0$,\hspace{5pt}
    \begin{enumerate}\vspace{-9pt}
        \item the input \( u(k) \) belongs to a known compact set \( \mathbb{U} \);\vspace{-9pt}
        \item the measurement noise is bounded, i.e., \( |w(k)| \leq \bar{w} \).\label{ass:boundednoise}
    \end{enumerate}
\end{ass}

We collect an input--output dataset \( \mathcal{D}_N = \{(u(k), y(k)), \, k = 0, \ldots, N-1\} \) from system \( \mathcal{S} \). 
These data are then used to select a suitable model structure within the candidate model family introduced in Section~\ref{subsec:candidatModelFamily} and to estimate the corresponding parameters, with the aim of obtaining a model that accurately represents the dynamical behavior of system \( \mathcal{S}\).

\subsection{Candidate model family} \label{subsec:candidatModelFamily}
We define a generic dynamical model as $\mathcal{M}(\eta,\theta)$, where $\mathcal{M}(\cdot\,,\cdot)$ represents the model family, $\eta \in H$ are the model hyperparameters that determine the model class, and $\theta \in \Omega$ are the model parameters that define any specific model within the class. The parameter vector $\theta$ must be learned directly from data, while the hyperparameters $\eta$ are usually selected manually, typically by gradually increasing the model complexity and choosing the configuration that provides the best trade-off between accuracy and parsimony.
 Here, $\Omega$ represents all possible parameterizations of the model class (i.e., $\theta\in\Omega$), and $H$ is the set of all possible model classes within the model family (i.e., $\eta\in H$. In this work we consider as candidate models, the ones in the family of NARXESNs, whose dynamics is described by the equations
\begin{subequations}\label{eq:NARXESN_candidateModel}
\renewcommand{\theequation}{\theparentequation.\alph{equation}}
\begin{align}
    \chi(k+1) &= \zeta\big(W_\chi\chi(k) + W_{\bm \varphi}{\bm \varphi}(k) + W_z z(k+1)\big), \label{eq:NARXESN_candidateReservoir} \\
    \hspace{-7.2pt}\mathcal{M}(\eta,\theta):\quad z(k+1) &= \theta^\top \psi(k), \label{eq:NARXESN_candidateOutput} \\
    y(k) &= z(k) + w(k), \label{eq:NARXESN_candidateMeasurement}
\end{align}
\end{subequations}
where $\chi\in\mathbb{R}^\nu$ is the state vector, $u\in\mathbb{R}$, $z\in\mathbb{R}$ and $y\in \mathbb{R}$ are the model input, noise-free output and measured output, respectively, whereas, $w\in\mathbb{R}$ is the measurement noise and $\zeta(\cdot)=[\,\zeta_1(\cdot),\, \ldots ,\,\zeta_\nu(\cdot)\,]^\top\in\mathbb{R}^\nu$ is a vector of functions $\zeta_i(\cdot):\mathbb{R}\rightarrow\mathbb{R}$. The vector $\psi(k)\in\mathbb{R}^{\nu+n_{\bm \varphi}}$ is defined as $\psi(k)=[\chi(k)^\top {\bm \varphi}(k)^\top]^\top$, where ${\bm \varphi}(k)=[\varphi_1 \ldots \varphi_{n_{\bm \varphi}}]^\top\in\mathbb{R}^{n_{\bm \varphi}}$ is the regressor vector that has as components $n_{\bm \varphi}$ elements drawn from the set
\begin{equation}\label{eq.candidate_regressors_set}
    \Phi(n_z,n_u)= \big\{z(k) , \ldots , z(k - n_z + 1) , u(k), \ldots , u(k - n_u + 1) \big\}.
\end{equation}
Matrices $W_\chi \in \mathbb{R}^{\nu \times \nu}$, $W_{\bm \varphi} \in \mathbb{R}^{\nu \times n_{\bm \varphi}}$ and $W_z \in \mathbb{R}^{\nu}$ are, according to ESN training procedure \cite{armenio_model_2019}, defined based on the hyperparameters. Indeed, the hyperparameters are collected in
\begin{equation}
\eta = \left\{ \Phi^\textrm{sel},\; \nu,\; n_{nl},\; \rho,\; k_z \right\},
\end{equation}
where $\Phi^\textrm{sel} \subseteq \Phi(n_z, n_u)$ denotes the set of selected regressors used to construct the vector ${\bm \varphi}(k)$, $\nu \in \mathbb{N}$ is the dimension of $\chi(k)$, $n_{nl} \in [0, \nu]$ is the number of nonlinear components in $\zeta(\cdot)$, $\rho \in (0,1)$ is the spectral radius of the matrix $W_\chi$ in \eqref{eq:NARXESN_candidateReservoir}, and $k_z \in \mathbb{R}$ is a scaling factor such that $W_z = k_z \tilde{W}_z$ for a fixed $\tilde{W}_z \in \mathbb{R}^{\nu}$.

On the other hand, $\theta=[W_{out_1}^\top\; W_{out_2}^\top]^\top$, with  $W_{out_1} \in \mathbb{R}^{\nu}$ and $W_{out_2} \in \mathbb{R}^{n_{\bm \varphi}}$ is the vector of free learnable parameters of the model\footnote{Formally, the dimension of $\theta$ depends on the hyperparameters $\eta$, since $W_{out_1}\!\in\mathbb{R}^{\nu}$ and $W_{out_2}\!\in\mathbb{R}^{n_{\bm\varphi}}$ with $n_{\bm\varphi}=|\Phi^{\mathrm{sel}}|$. For notational simplicity, this dependence is omitted throughout the paper.}.

For all models in family $\mathcal{M}(\eta,\theta)$, Assumption~\ref{ass:echostate_roperty} holds.

\begin{ass}\label{ass:echostate_roperty}
    The dynamic equation $$\chi(k+1)=\zeta\left(W_\chi\chi(k)\right)$$ enjoys the echo state property, i.e., the effect of the initial condition $\chi(0)$ asymptotically vanishes, see \cite{yildiz_re-visiting_2012}.
\end{ass}
The role of the Assumption \ref{ass:echostate_roperty}  in this work will be clarified later in Section \ref{sec:NARXESN_SM_setid}.
As discussed in \cite{yildiz_re-visiting_2012}, the echo state property can be conferred by setting $\bar{\sigma}(W_\chi)<1$, where $\bar{\sigma}(W_\chi)$ is the largest singular value of $W_\chi$. A less conservative condition on $W_\chi$ is provided by Theorem 4.1 in \cite{yildiz_re-visiting_2012}, namely the existence of a diagonal matrix $P>0$ such that $W_\chi^\top P W_\chi - P < 0$. A further relaxation of this condition is discussed in \cite{damico_2025_data}, where matrix $P>0$ is required to be diagonal only for the entries related to internal units with non linear activation function.

The identification procedure relies on the assumption that the system $\mathcal{S}$ belongs to the considered model family $\mathcal{M}(\cdot,\cdot)$. Formally, we assume the following.

\begin{ass}\label{ass:modelInConsideredFamily}
    Given the unknown dynamical system $\mathcal{S}$ and the family $\mathcal{M}(\cdot,\cdot)$, we assume that $\exists \,\eta^\circ \in H $ and $\theta^\circ \in \Theta$ such that $\mathcal{S} = \mathcal{M}(\eta^\circ, \theta^\circ)$.
\end{ass}

Under this assumption, the goal of model class selection is to find the hyperparameters $\eta^*$ such that $\mathcal{S} \in \mathcal{M}(\eta^*, \theta)$. Moreover, the parameter identification step will allow to select, among the models in set $\mathcal{M}(\eta^*,\theta)$, the best $\theta^*$ minimizing specific distance-based metrics, discussed in Section \ref{sec:NARXESN_SM_setid}.

\subsection{The general procedure}\label{subsec:generalproceduresketch}

In this section we sketch the general model class selection procedure that should be performed, whereas details on the different steps will be provided in the next sections. In particular, the following steps need to be carried out.

\begin{enumerate}[label=\alph*),ref=\alph*]
    \item \textbf{For all $\eta \in H$}:\label{step:a}
    \begin{enumerate}[label=\arabic*.,ref=a.\arabic*]
        \item We compute the set $\Theta\subseteq\Omega$ of free parameters $\theta$ compatible with the data and prior noise information using the set-membership approach outlined in Section \ref{sec:NARXESN_SM_setid}.\label{substep:1}
        \item We denote with $\mathcal{M}(\eta,\theta)$ for all $\theta \in \Theta$ as the ``optimal'' model set in class $\mathcal{M}(\eta,\cdot)$.\label{substep:2}
        \item Through a scenario-based sampling of $\theta$ in set $\Theta$ we simulate the possible trajectories of the models $\mathcal{M}(\eta,\theta)$ with $\theta\in\Theta$ and we compute $d_\mathcal{S}^*(\mathcal{M}(\eta,\cdot))$ as the minimum squared distance between the simulated outputs and the tube of possible output trajectories. The latter tube is obtained by jointly considering the output data and the a-priori assumptions on the measurement noise.\label{substep:3}
        \item Define the ``best" model in class $\mathcal{M}(\eta,\cdot)$ as $\mathcal{M}_*(\eta)=\mathcal{M}(\eta, \theta^*)$, where $\theta^* \in \Theta$ is the parametrization\footnote{Note that a distinct optimal parameter vector is obtained for each model class $\eta$, since the feasible set $\Theta$ and the corresponding minimizer
depend on $\eta$. To avoid overburdening the notation we do not explicitly write
$\theta^*(\eta)$, although this dependence is implicitly understood.} minimizing the set distance.\label{substep:4}
    \end{enumerate}
    \item \textbf{Compute the model class that minimizes $d^*_\mathcal{S}(\mathcal{M}_*(\eta))$}:
    \[
    \eta^* = \arg\min_{\eta\, \in\, H} d^*_\mathcal{S}(\mathcal{M}_*(\eta))
    \]\label{step:b}
    \item \textbf{Return $\mathcal{M}(\eta^*, \theta^*)$} as the identified optimal model.
\end{enumerate}

As discussed, the method adopts a set-membership identification approach in Step~\ref{substep:1}, under Assumption~\ref{ass:NARXESN_SM_assumptions}.\ref{ass:boundednoise} (that requires that the output noise is within known bounds), to determine an uncertain set of parameterizations $\Theta$, called the Feasible Parameter Set (FPS), compatible with system data. The ``suitability index'', i.e., the set-distance \(d_\mathcal{S}^*\), is used, e.g. in Step~\ref{substep:3}, to select the optimal parameter from the FPS (see Step~\ref{substep:4}) and the most suitable model class (see Step~\ref{step:b}). This performance index is obtained by measuring the quadratic distance between potential outputs within noise bounds and the closest simulated output.

Because the number of admissible model classes in $H$ increases combinatorially with the number of structural choices (regressors, states, activation functions, etc.), a full enumeration of $H$ is not computationally feasible. 
To avoid incurring in the curse of dimensionality, the procedure sketched above is replaced by an iterative heuristics described in Section \ref{sec:iterativeModelClassSelection}.

\section{Set Membership identification for NARXESN models}\label{sec:NARXESN_SM_setid}

In this section, we first assume that, for a given \(\eta \in H\), $\theta\in\Omega$ must be learned from data, being $\Omega$ an a-priori defined and aribtrairly large parameter set where $\theta$ can be found. All models in this class evolve according to the dynamics described in \eqref{eq:NARXESN_candidateModel}. The objective of this section is to describe in details steps~\ref{substep:1}--~\ref{substep:4} sketched in Section \ref{subsec:generalproceduresketch} aiming to determine the best model $\mathcal{M}_*(\eta)$ in this class. To do so, as decribed in Section~\ref{subsec:DefinitionoftheFeasibleParameterSet}, we first define a feasible parameter set (FPS) \(\Theta\), using a set-membership (SM) approach, which contains all parameter vectors in the given class that are consistent with the data generated by the system \(\mathcal{S}\) and the noise bound \(\bar{w}\) specified in Assumption~\ref{ass:NARXESN_SM_assumptions}. Based on this, $\mathcal{M}_*(\eta)$ will be selected as the model $\mathcal{M}(\eta;\theta)$ for $\theta\in\Theta$ which minimizes the set-distance as defined in Section~\ref{subsec:defOfOptimalModelInTheClass} to evaluate model consistency with the data.

\subsection{Definition of the feasible parameter set}\label{subsec:DefinitionoftheFeasibleParameterSet}

Set Membership (SM) identification, introduced by Witsenhausen \cite{witsenhausen_sets_1968} and Schweppe \cite{schweppe_recursive_1968}, provides a framework for estimating unknown parameters from data. Unlike traditional stochastic approaches, SM assumes that uncertainty is bounded within a known range, enabling the identification of the FPS, containing all parameter values consistent with available data and assumed the noise bounds \cite{milanese_optimal_1996}. Although SM identification can be challenging for general-type nonlinear systems \cite{milanese_set_2004}, models with a linear-in-the-parameters (LITP) structure \eqref{eq:NARXESN_candidateOutput}, allow for an efficient application of the SM approach.

Since only the output measurements \( y(k) \) are available instead of the nominal output \( z(k) \), the regressor vector must be reconstructed from noisy data. We define the estimated regressor vector \( \hat{{\bm \varphi}}(k) \) by replacing each nominal output \( z(\cdot) \) in the set \( \Phi^\textrm{sel} \subseteq \Phi(n_z, n_u) \) with its noisy counterpart \( y(\cdot) \), i.e.,
\[
\hat{{\bm \varphi}}(k) = {\bm \varphi}(k)\big|_{z(\cdot)\rightarrow y(\cdot)}\,,
\]
where \( {\bm \varphi}(k) = [\varphi_1(k), \ldots, \varphi_{n_{\bm \varphi}}(k)]^\top \) and each \( \varphi_i(k) \in \Phi^\textrm{sel} \) is a delayed signal \( z(k - \ell) \) or \( u(k - \ell) \).

The internal reservoir state \( \chi(k) \), which is not directly measurable, in view of Assumption \ref{ass:echostate_roperty} can be estimated recursively as:\begin{equation}
    \hat{\chi}(k+1) = \zeta\left(W_\chi\hat{\chi}(k)+W_{\bm \varphi}\hat{{\bm \varphi}}(k)+W_z y(k+1)\right).
\end{equation}

The noisy vector \( \hat{\psi}(k) \) is constructed as
\begin{equation*}
    \hat{\psi}(k) =
    \begin{bmatrix}
        \hat{\chi}(k) \\
        \hat{{\bm \varphi}}(k)
    \end{bmatrix}.
\end{equation*}

Substituting equation \eqref{eq:NARXESN_output_LITP} into the output relationship \eqref{eq:NARXESN_measurement}, the latter can be rewritten as
\begin{equation}\label{eq:noisy_equation_sys}
    y(k+1) = \theta^{\circ^\top}\psi^\circ(k) + w(k+1).
\end{equation}
Moreover, the estimation error obtained substituting $\hat{\psi}(k)$ in \eqref{eq:NARXESN_candidateOutput} is \[\varepsilon(k)=\theta^{\circ^\top}\psi^\circ(k)-\theta^{\top}\hat\psi(k).\] Isolating term $\theta^{\circ^\top}\psi^\circ(k)=\theta^{\top}\hat\psi(k)+\varepsilon(k)$ and plugging it into \eqref{eq:noisy_equation_sys} we obtain\begin{equation}\label{eq:eq:noisy_equation_model}
y(k + 1) = \theta^{\top} \hat{\psi}(k) + \varepsilon(k) + w(k + 1).
\end{equation}

Given a dataset of \( N \) output/regressor data pairs \( (y(k+1), \hat{\psi}(k)) \), the Feasible Parameter Set (FPS) is defined as the set of all parameter values consistent with the available information and can be computed, in view of LITP structure of \eqref{eq:eq:noisy_equation_model}, using the procedure in Algorithm \ref{alg:SM_FPSid_gen} inspired from \cite{damico_2025_data}.

\begin{algorithm}[H]
    \caption{Estimation of the FPS via Set Membership}
    \small
    \label{alg:SM_FPSid_gen}
    \begin{algorithmic}[1]
    \REQUIRE Noise amplitude upper bound $\bar w$, compact set $\Omega$
    \ENSURE Feasible parameter set $\Theta$
        \STATE Collect output/regressor data pairs \( (\,y(k+1),\, \hat{\psi}(k)\,) \) for all \( k=0,\ldots,N-1 \).
        \STATE Obtain an underestimate \( \underline{\lambda} \) of the prediction error bound by solving:
        \begin{equation}\label{eq:lambda_underbarEstimate}
\begin{aligned}
    \underline{\lambda} = &\min_{\theta \in \Omega, \lambda \geq 0} \quad \lambda \\
    &\quad \text{s.t. } \quad
    \big|y(k+1)-\theta^\top \hat{\psi}(k)\big| \leq \lambda + \bar{w}, 
    \quad \forall k
\end{aligned}
\end{equation}
        \STATE Compute an estimate of the disturbance bound, by slightly inflating $\lambda$ with a parameter $\alpha$ as
        \begin{equation*}
            \hat{\bar{\varepsilon}}(\alpha) = \alpha \underline{\lambda}, \quad \alpha \geq 1
        \end{equation*}
        \STATE Define the FPS as:
        \begin{equation}\label{eq:FPS_definition}
            \Theta(\alpha) := \left\{\theta \in \Omega : \big|y(k+1) - \theta^\top \hat{\psi}(k)\big| \leq \hat{\bar{\varepsilon}}(\alpha) + \bar{w}, \quad \forall k \,\right\}
        \end{equation}
        \STATE $\Theta\gets\Theta(\alpha)$
        \RETURN $\Theta$
    \end{algorithmic}
\end{algorithm}

The estimate \( \underline{\lambda} \) in \eqref{eq:lambda_underbarEstimate} represents the minimal achievable error bound, which is then scaled by a factor \( \alpha \geq 1 \) in \eqref{eq:FPS_definition} to account for the limited dataset. The parameter \( \alpha \) can be estimated using:

\begin{equation}\label{eq:alphaEstimate}
    \alpha = \max_{k=0,\ldots,N-1}\frac{\big|y(k+1)-\hat{\theta}^{\top}\hat{\psi}(k)\big|-\bar{w}}{\underline{\lambda}}
\end{equation}
where \( \hat{\theta} \) is an initial parameter estimate obtained via least squares (LS) optimization. 

The FPS \( \Theta(\alpha) \) provides a guaranteed bound on the true parameter values, offering a robust region for validation and further analysis.

\subsection{Definition of the optimal model in the set} \label{subsec:defOfOptimalModelInTheClass}

Before to define the procedure for estimating the optimal model in the set, we need to formally introduce the definition of set-distance.


Consider a model $\mathcal{M}\left(\eta,\theta\right)$ of class $\eta$ and with parameter $\theta$, the set distance metric is defined as the sum of distances between the model's simulated output $y_\theta(k)$ with $\theta$ and the tube of all possible outputs trajectories:
\begin{equation}\label{eq:distances_from_tube}
    d(\mathcal{M}(\eta,\theta))= \sum_{k=N_w,\ldots,N-1} \text{dist}(y_\theta(k), \mathcal{Y}(k))
\end{equation}

where \( \mathcal{Y}(k) = \{ \tilde{y}(k) : | \tilde{y}(k) - y(k) | \leq \bar{w} \}\) represents the set of all possible outputs within noise bounds, \( N_w \ge 0 \) is a suitable washout time such that the transient effects of the neural network states have vanished, as guaranteed by Assumption~\ref{ass:echostate_roperty} and \(\text{dist}(v, \mathcal{O}) = \min_{x \in \mathcal{O}} \| v - x \|^2\) is the point-to-set distance.

The objective is to find the best parametrization $\theta^*$ within the feasible parameter set (FPS) \( \Theta \), i.e., 
\[
\theta^* = \arg\min_{\theta \in \Theta} d(\mathcal{M}(\eta,\theta)).
\]

Due to the impracticality of evaluating all possible parameterizations within the FPS, we propose an approximate method based on the scenario approach \cite{campi_scenario_2008}. This involves sampling points $\theta_i$, for $i=1,\ldots,\mathcal{N}_s$ from the FPS using a probability distribution to ensure the sample is representative of the entire FPS. At this point we select the optimal parametrization $\theta^*$ as $\theta_{i^*}$, where 
\begin{equation}\label{eq:best_parametr_index}
    i^*=\arg\min_{i=1,\ldots,\mathcal{N}_s} d(\mathcal{M}(\eta,\theta_i)).
\end{equation}
In turn, we denote as $d_\mathcal{S}^*(\mathcal{M}(\eta,\cdot))=d(\mathcal{M}(\eta,\theta_{i^*}))$.

The number of scenarios \( \mathcal{N}_s \) is determined to ensure that the probability of finding a new scenario with a smaller distance is below a specified threshold. This, according to \cite{damico_2025_data}, is achieved by selecting \( \mathcal{N}_s \) to satisfy the following inequality:
\begin{equation}\label{eq:scenarios_ineq}
\mathcal{N}_s \geq \frac{\log(\beta)}{\log(1-\epsilon)},
\end{equation}
where \( \epsilon \in (0,1) \) is the acceptable probability that a new scenario would have a smaller distance than the current minimum, and \( \beta \in (0,1) \) is the confidence level that ensures with probability \( 1 - \beta \) that the sampled scenarios adequately represent the FPS. This ensures, with similar arguments to \cite{damico_data-based_2023}, that the number of scenarios \( \mathcal{N}_s \) is chosen to provide a high confidence that the sampled scenarios adequately represent the FPS.

To draw representative samples from the FPS, we estimate the probability distribution \( \mathbb{P}_\theta \) over the FPS using the mean and covariance matrix derived from Prediction Error Minimization (PEM) methods. For a Linear-In-the-Parameters (LITP) model, we can estimate \( \mathbb{P}_\theta \) adopting as its mean and covariance, the least squares (LS) estimator and the covariance of the latter (see Section 5 of \cite{campi_virtual_2000} and references therein for further details).

The procedure for estimating the optimal parameter vector $\theta^*$ within the feasible parameter set described above is summarized in Algorithm~\ref{alg:parametersIdentification}.

\begin{algorithm}[H]
\small
\caption{Parameters identification}\label{alg:parametersIdentification}
\begin{algorithmic}[1]
\REQUIRE Model class $\mathcal{M}(\eta\,,\cdot)$
\ENSURE Optimal Model $\mathcal{M}(\eta\,,\theta^*)$ and set-distance $d_\mathcal{S}^*(\mathcal{M}(\eta\,,\cdot))$
\STATE Compute FPS $\Theta$ using Algorithm \ref{alg:SM_FPSid_gen}
\STATE Define a probability distribution $\mathbb{P}_\theta$ over $\Theta$ as in \cite[Section 5]{campi_virtual_2000}
\STATE Select a finite number of scenarios $\mathcal{N}_\mathcal{S}$ such that \eqref{eq:scenarios_ineq} holds
\FOR{$i=1,\ldots,\mathcal{N}_\mathcal{S}$}
    \STATE Draw $\theta_i\in\Theta$ with probability $\mathbb{P}_\theta$
    \STATE Simulate output $y_{\theta_i}(k)$ for $k=N_w,\ldots,N-1$ of model $\mathcal{M}(\eta\,,\theta_i)$
    \STATE Compute $d(\mathcal{M}(\eta\,,\theta_i))$ as in \eqref{eq:distances_from_tube}
    \STATE Compute index $i^*$  minimizing $d(\mathcal{M}(\eta\,,\theta_i))$ as in \eqref{eq:best_parametr_index}
\ENDFOR
\STATE Choose optimal parametrization $\theta^*$ of the model: $\theta^*\gets\theta_{i*}$
\STATE Define set-distance of model class $\mathcal{M}(\eta\,,\cdot)$ as $d^*_\mathcal{S}(\eta\,;\cdot)\gets d(\mathcal{\eta;\theta^*})$
\RETURN $\mathcal{M}(\eta\,;\theta^*)\,,\,\, d_\mathcal{S}^*(\mathcal{M}(\eta\,,\cdot))$
\end{algorithmic}
\end{algorithm}

\section{Iterative Model Class Selection}\label{sec:iterativeModelClassSelection}
\begin{figure}[h!]
\centering
\includegraphics[width=0.525\textwidth]{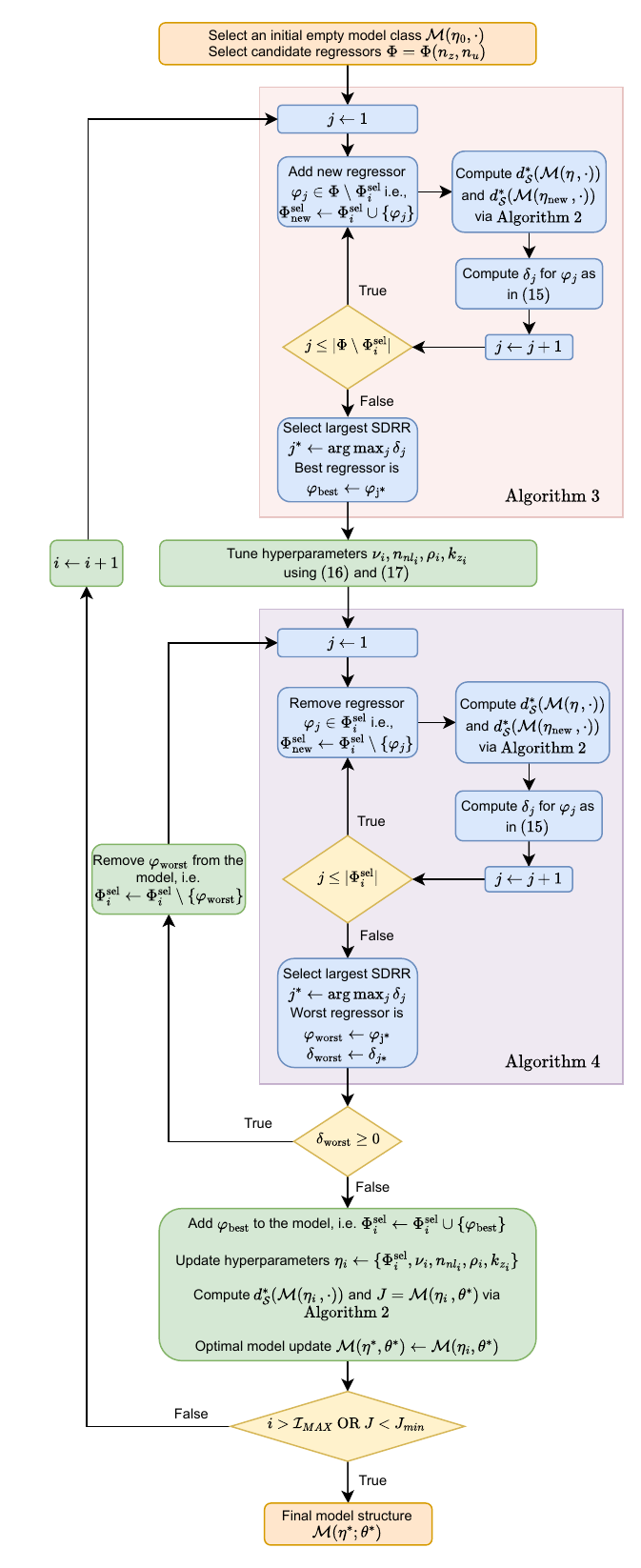} 
\caption{Flow chart of Algorithm~\ref{alg:NARXESNmodelStructureSelection}.}
\label{fig:FluxDiagram}
\end{figure}
The proposed iterative heuristics for model class selection (i.e., step \ref{step:b} in Section \ref{subsec:generalproceduresketch}) is sketched in the flow diagram in Figure \ref{fig:FluxDiagram}. At each step we add significant features and remove less significant ones based on the set-distance metric. The outer loop handles feature inclusion/exclusion, while the inner loop provides parameter learning using set-membership identification \cite{milanese_optimal_1996} to define the FPS and determine the best parameterization. Moreover, at each iteration a tuning step of the so-called ``numerical" hyperparameters, namely $\nu,n_{nl},\rho$ and $k_z$,  is performed.

This section is organized as follows: Section \ref{subsec:RegressorsSelectionandPruning} describes how regressors are selected and pruned using the SDRR criterion; Section \ref{subsec:NumericalHyperparametersTuning} details the iterative tuning of discrete and continuous hyperparameters; finally, Section \ref{subsec:SummaryoftheMethod} presents a summary of the full method, including the strategy for model selection based on multiple initializations.

\subsection{Regressors Selection and Pruning}\label{subsec:RegressorsSelectionandPruning}

To define the best regressor collection $\Phi^{\textrm{sel},*}\subseteq\Phi(n_z,n_u)$, we adopt a forward selection approach, recursively adding the best regressor $\varphi_\textrm{best}\in\Phi(n_z,n_u)\setminus\Phi^\textrm{sel}$ to the model based on the set-distance reduction ratio (SDRR). The latter measures the relevance of a candidate regressor $\varphi_j$ in terms of set-distance metric. Consider two model classes $\mathcal{M}(\eta\,,\cdot)$ and $\mathcal{M}(\eta_\textrm{new}\,,\cdot)$, where the only difference between these classes is the fact that $\mathcal{M}(\eta_\textrm{new}\,,\cdot)$ includes the additional regressor $\varphi_j$. The SDRR is defined as 
\begin{equation}\label{eq:sdrr_definition}
    \delta_j = \frac{d^*_\mathcal{S}(\mathcal{M}(\eta\,,\cdot)) - d^*_\mathcal{S}(\mathcal{M}(\eta_\textrm{new}\,,\cdot))}{\frac{1}{N} \sum_{k=0}^{N-1} y^2(k)}.
\end{equation}

To further refine the model, inspired by \cite{piroddi_identification_2003}, we employ pruning to remove terms that minimally enhance the model's quality, i.e., that corresponds to a negligible value of $\delta_j$.

\medskip
Summarizing, the forward selection procedure starts with an initial model class $\mathcal{M}(\eta_{0},\cdot)$. Then, we iteratively add the regressor that maximizes the SDRR to the model at each iteration. After the initial iterations, we can also remove, i.e., prune, (if this improves the model) the regressor $\varphi_\textrm{worst}\in\Phi^\textrm{sel}$ that has the least impact on the set-distance.
The implementation of the forward selection and pruning mechanisms is summarized in Algorithms~\ref{alg:bestRegressorSelection} and~\ref{alg:worstRegressorSelection}, which detail the procedures used to identify the most and least relevant regressors according to the SDRR criterion.

\begin{algorithm}[H]
\small
\caption{Best regressor selection}\label{alg:bestRegressorSelection}
\begin{algorithmic}[1]
\REQUIRE candidate regressors $\Phi(n_z,n_u)$, model class $\mathcal{M}(\eta\,,\cdot)$ with $$\eta=\{\Phi^\textrm{sel},\nu,n_{nl},\rho,k_z\}$$
\ENSURE Best candidate regressor $\varphi_\textrm{best}$
\FORALL{$\varphi_j\in\Phi(n_z,n_u)\setminus\Phi^{\textrm{sel}}_i$}
    \STATE{Build model class $\mathcal{M}(\eta_\textrm{new},\cdot)$ including new regressor $\varphi_j$, i.e., $$\Phi^\textrm{sel}_\textrm{new}\gets \Phi^\textrm{sel} \cup  \{\varphi_j\},\quad\eta_\textrm{new}\gets\{\Phi^\textrm{sel}_\textrm{new},\nu,n_{nl},\rho,k_z\}$$}
    \STATE Compute $d_\mathcal{S}^*(\mathcal{M}(\eta\,,\cdot))$ and $d_\mathcal{S}^*(\mathcal{M}(\eta_\textrm{new},\cdot))$ using Algorithm \ref{alg:parametersIdentification}
    \STATE Compute SDRR $\delta_j$ for term $\varphi_j$ as in \eqref{eq:sdrr_definition}
\ENDFOR
\STATE Choose $\varphi_\textrm{best}$ as the regressor whose inclusion in the model provides the largest SDRR, i.e.,
$$\varphi_\textrm{best}\gets\varphi_{j^*}, \quad \textrm{where }j^{\,*}\gets \arg\max_j \delta_j$$
\RETURN $\varphi_\textrm{best}$
\end{algorithmic}
\end{algorithm}

\begin{algorithm}[H]
\small
\caption{Worst regressor selection}\label{alg:worstRegressorSelection}
\begin{algorithmic}[1]
\REQUIRE Model class $\mathcal{M}(\eta\,,\cdot)$ with $\eta=\{\Phi^\textrm{sel},\nu,n_{nl},\rho,k_z\}$
\ENSURE Worst model regressor $\varphi_\textrm{worst}$ and corresponding SDRR $\delta_\textrm{worst}$
\FORALL{$\varphi_j\in\Phi^\textrm{sel}$}
    \STATE{Build model class $\mathcal{M}(\eta_\textrm{new},\cdot)$ excluding regressor $\varphi_j$, i.e., $$\Phi^\textrm{sel}_\textrm{new}\gets \Phi^\textrm{sel} \setminus  \{\varphi_j\},\quad\eta_\textrm{new}\gets\{\Phi^\textrm{sel}_\textrm{new},\nu,n_{nl},\rho,k_z\}$$}
    \STATE Compute $d_\mathcal{S}^*(\mathcal{M}(\eta\,,\cdot))$ and $d_\mathcal{S}^*(\mathcal{M}(\eta_\textrm{new},\cdot))$ using Algorithm \ref{alg:parametersIdentification}
    \STATE Compute SDRR $\delta_j$ for term $\varphi_j$ as in \eqref{eq:sdrr_definition} 
\ENDFOR
\STATE Choose $\varphi_\textrm{worst}$ as the regressor whose exclusion from the model provides the largest SDRR, i.e.,
$$\varphi_\textrm{best}\gets\varphi_{j^*}, \quad \textrm{where }j^{\,*}\gets \arg\max_j \delta_j$$
\STATE Select $\delta_\textrm{worst}\gets\delta_{j^*}$
\RETURN $\varphi_\textrm{worst}$, $\delta_\textrm{worst}$
\end{algorithmic}
\end{algorithm}

\subsection{Numerical Hyperparameters Tuning}\label{subsec:NumericalHyperparametersTuning}
The tuning of the numerical hyperparameters is performed iteratively. At the
beginning of iteration $i$, the regressor set is denoted by
$\Phi_i^{\mathrm{sel}}$, because the best regressor to be added has just been
identified according to the procedure described in Algorithm~\ref{alg:bestRegressorSelection} and included
in the model. Given this updated structure, the current hyperparameter
configuration is
\[
\eta_{i-1}=\{\Phi_i^{\mathrm{sel}},\, \nu_{i-1},\, n_{nl_{i-1}},\, \rho_{i-1},\, k_{z_{i-1}}\},
\]
and the numerical hyperparameters are refined in a round-robin fashion while keeping the
others fixed.
The updated numerical hyperparameters obtained from the tuning step are denoted
by $\nu_i$, $n_{nl_i}$, $\rho_i$ and $k_{z_i}$, to indicate that they correspond
to the configuration associated with the regressor set $\Phi_i^{\mathrm{sel}}$.

 For the integer hyperparameters $\nu$ and $n_{nl}$, the set-distance is evaluated at neighboring integer values, i.e., within the ranges $N_\chi = [\nu_{i-1} - 1,\, \nu_{i-1} + 1]$ and $N_{nl} = [n_{nl_{i-1}} - 1,\, n_{nl_{i-1}} + 1]$, respectively. The values yielding the smallest set-distance are selected. Formally,
\begin{subequations}\label{eq:discrete_numHyperpTuning}
\renewcommand{\theequation}{\theparentequation.\alph{equation}}
\begin{align}
\nu_i &= \arg\min_{\nu\in N_\chi} d_\mathcal{S}^*(\mathcal{M}(\eta_{i-1}(\nu),\cdot)), \\
n_{nl_i} &= \arg\min_{n_{nl}\in N_{nl}} d_\mathcal{S}^*(\mathcal{M}(\eta_{i-1}(n_{nl}),\cdot)),
\end{align}
\end{subequations}
where $d_\mathcal{S}^*(\mathcal{M}(\eta_{i-1}(x)),\cdot)$ denotes the set-distance metric of the model class $\mathcal{M}(\eta_{i-1}(x), \cdot)$, with only the hyperparameter $x$ (either $\nu$ or $n_{nl}$) being varied.

On the other hand, for the continuous hyperparameters $\rho$ and $k_z$, the Newton's method is applied to minimize the set-distance, according to the update rules
\begin{subequations}\label{eq:continous_numHyperpTuning}
\renewcommand{\theequation}{\theparentequation.\alph{equation}}
\begin{align}
    \rho_i &= \rho_{i-1} - \frac{\frac{\partial}{\partial\rho} d^*_\mathcal{S}(\mathcal{M}(\eta_{i-1}(\rho),\cdot))}{\frac{\partial^2}{\partial\rho^2} d^*_\mathcal{S}(\mathcal{M}(\eta_{i-1}(\rho),\cdot))}, \\
    k_{z_i} &= k_{z_{i-1}} - \frac{\frac{\partial}{\partial k_z} d^*_\mathcal{S}(\mathcal{M}(\eta_{i-1}(k_z),\cdot))}{\frac{\partial^2}{\partial k_z^2} d^*_\mathcal{S}(\mathcal{M}(\eta_{i-1}(k_z),\cdot))}.
\end{align}
\end{subequations}
Since the analytical form of the set-distance derivatives is not available, the required first- and second-order derivatives in \eqref{eq:continous_numHyperpTuning} are numerically approximated using finite differences (see \cite[Chapter 8]{nocedal2006numerical}), i.e., by perturbing the corresponding hyperparameter and evaluating the resulting variation of $d^*_\mathcal{S}(\cdot)$. This allows the application of Newton's method without the need for explicit gradient expressions.

\subsection{Summary of the Method}\label{subsec:SummaryoftheMethod}
Given the initial hyperparameters $\eta_0=\{\Phi^\textrm{sel}_{0}, \nu_0, n_{nl_0}, \rho_0, k_{z_0}\}$, the overall identification and structure–selection procedure is summarized in Algorithm~\ref{alg:NARXESNmodelStructureSelection}. 
\begin{algorithm}[h]
\small
\caption{NARXESN model structure selection}\label{alg:NARXESNmodelStructureSelection}
\begin{algorithmic}[1]
\REQUIRE Candidate regressors $\Phi(n_z,n_u)$
\ENSURE Final model $\mathcal{M}(\eta^*;\theta^*)$
\STATE \textbf{Initialize:} Model class $\mathcal{M}(\eta_0;\cdot)$ with $\eta_0 = \{\varnothing, \nu_0, n_{nl_0}, \rho_0, k_{z_0}\}$
\STATE Define stopping criteria: set-distance tolerance $J_{min}$ and max iterations $\mathcal{I}_{MAX}$
\STATE Find best initial regressor $\varphi_{\text{best}}\in\Phi(n_z,n_u)$ using Algorithm \ref{alg:bestRegressorSelection}
\STATE Set iteration counter $i \gets0$
\WHILE{$J \geq J_{min}$ \AND $i \leq \mathcal{I}_{MAX}$}
   \STATE $i \gets i + 1$
   \STATE Add $\varphi_{\text{best}}$ to model, i.e., $\Phi^\textrm{sel}_i \gets \Phi^\textrm{sel}_{i-1} \cup \{\varphi_{\text{best}}\}$
   \STATE Tune $ \nu_i, n_{nl_i}, \rho_i, k_{z_i}$ using \eqref{eq:discrete_numHyperpTuning} and \eqref{eq:continous_numHyperpTuning}
   \STATE Find new $\varphi_{\text{best}}\in\Phi(n_z,n_u)\setminus\Phi^\textrm{sel}_i$ for next step using Algorithm \ref{alg:bestRegressorSelection}
   \IF{$i \geq 2$}
   \REPEAT
       \STATE Find $\varphi_\textrm{worst}\in\Phi^\textrm{sel}_i$ and related $\delta_\textrm{worst}$ using Algorithm \ref{alg:worstRegressorSelection}
       \STATE Prune $\varphi_\textrm{worst}$ from the model, i.e., $\Phi^\textrm{sel}_i \gets \Phi^\textrm{sel}_i \setminus \{\varphi_\textrm{worst}\}$
    \UNTIL{$\delta_\textrm{worst}\geq 0$}
   \ENDIF
   \STATE Compute $d^*_\mathcal{S}(\mathcal{M}(\eta_i;\cdot))$ and $\mathcal{M}(\eta_i;\theta^*)$ using Algorithm \ref{alg:parametersIdentification}
   \STATE Assign model class hyperparameters $\eta_i\gets\{\Phi^\textrm{sel}_i,\nu_i,n_{nl_i},\rho_i,k_{z_i}\}$
   \STATE Update set-distance for class $\mathcal{M}(\eta_i;\cdot)$, i.e., $\, J \gets d^*_\mathcal{S}(\mathcal{M}(\eta_i;\cdot))$
   \STATE Update optimal model: $\mathcal{M}(\eta^*;\theta^*)\gets\mathcal{M}(\eta_i;\theta^*)$
\ENDWHILE
\RETURN $\mathcal{M}(\eta^*, \theta^*)$
\end{algorithmic}
\end{algorithm}
The latter integrates the individual previously described phases, i.e., 
the parameter identification step (Algorithm~\ref{alg:parametersIdentification}), 
the regressor selection and pruning strategies (Algorithms~\ref{alg:bestRegressorSelection} and~\ref{alg:worstRegressorSelection}), 
and the hyperparameter tuning rules introduced in \eqref{eq:discrete_numHyperpTuning}–\eqref{eq:continous_numHyperpTuning}.

The method iteratively enhances the model structure, continuously seeking an optimal compromise between accuracy and model simplicity.

Due to the sensitivity of NARXESN models to hyperparameter initialization, 
the structure selection procedure is executed multiple times starting from 
different randomly generated initial hyperparameter vectors. Let
$
\eta_0^{\,l}$, with $l = 1,\ldots,\mathcal{N}_\text{init}$
denote the \(l\)-th random initialization. Each initialization is used as a 
starting point for Algorithm~\ref{alg:NARXESNmodelStructureSelection}, 
which returns a candidate optimal model
\[
\mathcal{M}\big(\eta^{*,\,l},\,\theta^{*,\,l}\big).
\]
Among the \(\mathcal{N}_\text{init}\) resulting models, the final selected model
is the one achieving the smallest set-distance:
\[
l^* = \arg\min_{\,l=1,\ldots,\mathcal{N}_\text{init}}
d^*_\mathcal{S}\!\left(\mathcal{M}\big(\eta^{*,\,l},\theta^{*,\,l}\big)\right).
\]
The final model is therefore chosen as
\[
\mathcal{M}(\eta^{*,\,l^*},\theta^{*,\,l^*}),
\]
i.e., the one obtained from the initialization that yields the best consistency 
with the data according to the proposed set-distance criterion.

\section{Simulation Results}\label{sec:SimulationResults}

In this section, we illustrate the results obtained by applying the proposed method to data generated by a simulated system belonging to the NARXESN model family.

\subsection{Simulated NARXESN System}
\begin{figure}[h]
    \centering
    \includegraphics[width=0.8\textwidth]{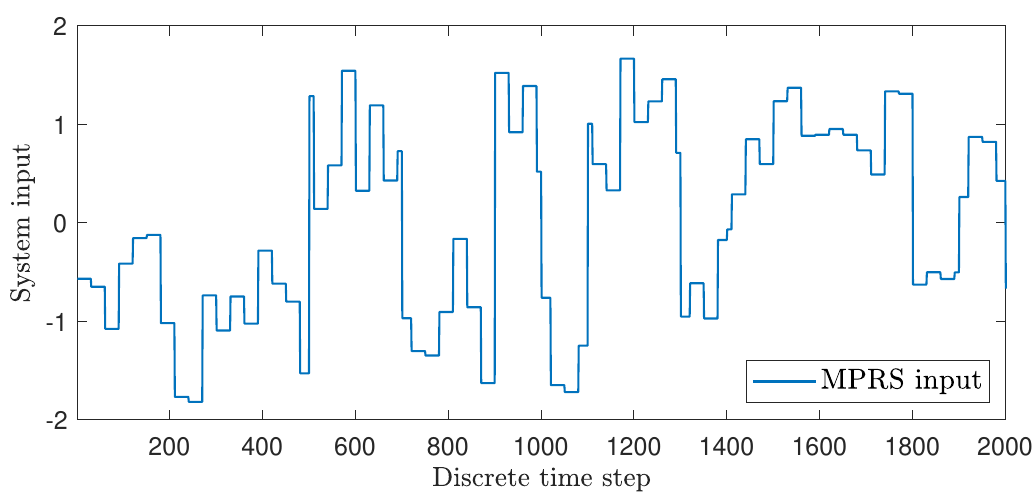} 
    \caption{MPRS input signal applied to the simulated NARXESN system.}
    \label{fig:NARXESN_MPRS_input}
\end{figure}

The system under consideration is a NARXESN model \eqref{eq:NARXESN_model} where $\chi^\circ(k) \in \mathbb{R}^{\nu^\circ}$ represents the state vector with $\nu^\circ = 6$ neurons, nominal parameters are
\[
W_{out_1}^\circ = [\,-0.0761,\, 0.1396,\, -0.0404,\, 0.0161,\, -0.0466,\, -0.2883\,],
\]
\[W_{out_1}^\circ = [\,0.2469,\, 0.2267,\, 0.3151,\, 0.2174],\]
${\bm \varphi}^\circ(k)$ is the regressor vector:
\[
{\bm \varphi}^\circ(k) = \begin{bmatrix} z(k) & z(k-1) & u(k) & u(k-1) \end{bmatrix}^\top,
\]
and $\zeta^\circ(\cdot)$ is the activation function vector with $n^o_{nl}=3$ nonlinear components:
\[
\zeta^\circ(\cdot) = \begin{bmatrix}
    \tanh(\cdot) & \tanh(\cdot) & \tanh(\cdot) & \text{id}(\cdot) & \text{id}(\cdot) & \text{id}(\cdot)
\end{bmatrix}^\top.
\]
The model parameters and hyperparameters are known, where the spectral radius is set to $\rho^\circ = 0.45$ and the feedback scaling factor is $k_z^\circ = 1$. The system is excited with a Multilevel Pseudo-Random Signal (MPRS) input $u(k)$ shown in Figure~\ref{fig:NARXESN_MPRS_input}, whereas the output is affected by uniform noise $w(k) \sim \mathcal{U}(-0.05, 0.05)$.

\subsection{Experiment and Results}\label{subsec:NARXESN_Results}

The structure selection algorithm was configured with a noise amplitude upper bound consistent with the noise realization, i.e., $\bar{w} = 0.05$ and applied to identify the correct model structure. The candidate regressors were selected from the set:
\[
\Phi = \{ z(k),\, z(k-1),\, z(k-2),\, u(k),\, u(k-1),\, u(k-2) \},
\]
and where the hyperparameters vary within the following ranges:
\[
\nu \in [1, 10], \quad n_{nl} \in [0, \nu], \quad \rho \in (0, 1), \quad k_z \in (0.5, 1.5).
\]

The algorithm iteratively refined the structure through selection, pruning, and hyperparameter tuning. Table~\ref{table:NARXESN_MSS_scenario12} reports the key iterations for the scenario where the correct structure was eventually identified. The final model identified has indeed the following structure:
\renewcommand{\arraystretch}{1.1}

\begin{table*}[t]
    \centering
    \caption{Iterative evolution of the NARXESN hyperparameters of scenario $s=12$.}
    \label{table:NARXESN_MSS_scenario12}

    \newcolumntype{M}[1]{>{\centering\arraybackslash}m{#1}}

    \resizebox{\textwidth}{!}{%
    \begin{tabular}{
        M{0.7cm}
        M{0.7cm}
        M{0.7cm}
        M{1.3cm}
        M{1.3cm}
        M{2.3cm}
        M{2.0cm}
        M{2.3cm}
    }
    \toprule
    \textbf{Iter.} & $\bm{\nu}$ & $\bm{n_{nl}}$ & $\bm{\rho}$ & $\bm{k_z}$ &
    \textbf{Add/Prune} & \textbf{Regressor} & $\bm{d_\mathcal{S}^*}$ \\
    \midrule

    \rowcolor{gray!10}
    $\bm{1}$ & 5 & 0 & 0.3114 & 0.95269 & $+$ & $u(k)$ &
    $9.3318 \times 10^{-2}$ \\

    $\bm 2$ & 5 & 1 & 0.3097 & 0.95312 & $+$ & $z(k)$ &
    $2.309 \times 10^{-3}$ \\

    \rowcolor{gray!10}
    $\bm 3$ & 6 & 2 & 0.30907 & 0.95055 &
    \shortstack{$-$\\$+$} &
    \shortstack{$z(k)$\\$z(k-1)$} &
    $1.6351 \times 10^{-1}$ \\

    $\bm 4$ & 6 & 1 & 0.3455 & 0.95144 &
    \shortstack{$-$\\$+$} &
    \shortstack{$z(k-1)$\\$z(k)$} &
    $6.2681 \times 10^{-4}$ \\

    \rowcolor{gray!10}
    $\bm 5$ & 6 & 1 & 0.35901 & 0.96923 & $+$ & $z(k-1)$ &
    $6.7799 \times 10^{-2}$ \\

    $\bm 6$ & 5 & 2 & 0.40744 & 1.0213 &
    \shortstack{$-$\\$+$} &
    \shortstack{$z(k-1)$\\$z(k-2)$} &
    $5.1172 \times 10^{-3}$ \\

    \rowcolor{gray!10}
    $\bm 7$ & 5 & 3 & 0.40194 & 1.0192 &
    \shortstack{$-$\\$+$} &
    \shortstack{$z(k-2)$\\$z(k-1)$} &
    $5.0776 \times 10^{-4}$ \\

    $\bm 8$ & 5 & 2 & 0.40419 & 1.017 & $+$ & $z(k-2)$ &
    $6.1885 \times 10^{-4}$ \\

    \rowcolor{gray!10}
    $\bm 9$ & 5 & 1 & 0.42528 & 1.006 & $+$ & $u(k-2)$ &
    $1.5783 \times 10^{-3}$ \\

    $\bm{10}$ & 5 & 2 & 0.42731 & 1.013 &
    \shortstack{$-$\\$-$\\$+$} &
    \shortstack{$u(k-2)$\\$z(k-2)$\\$u(k-1)$} &
    $1.4148 \times 10^{-3}$ \\

    \rowcolor{gray!10}
    $\bm{11}$ & 6 & 3 & 0.42359 & 1.018 & $+$ & $u(k-2)$ &
    $4.3761 \times 10^{-4}$ \\

    $\bm{12}$ & 6 & 3 & 0.45354 & 1.0178 & $-$ & $u(k-2)$ &
    $9.736 \times 10^{-7}$ \\
    \bottomrule
    \end{tabular}
    }
\end{table*}

\begin{figure}[H]
    \centering
    \includegraphics[width=0.8\textwidth]{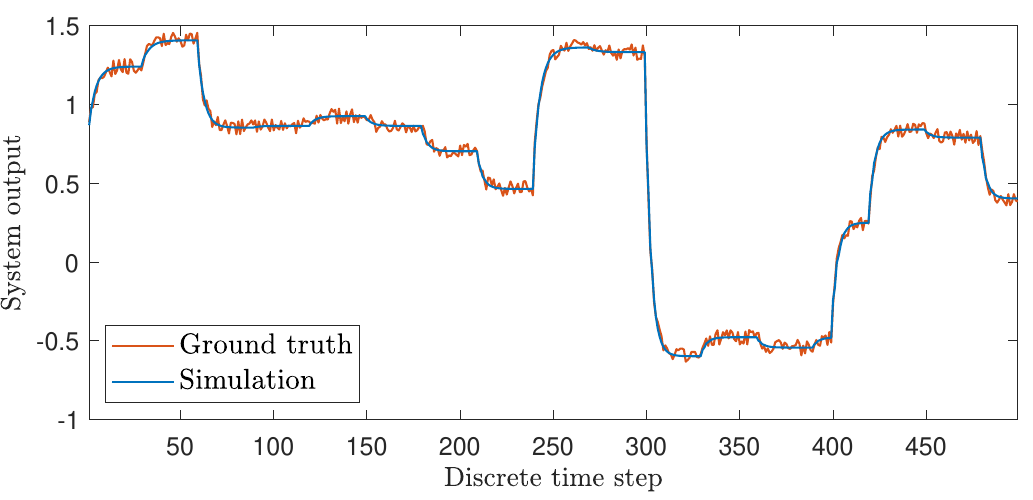}
    \vspace{-0.2cm}
    \caption{Simulation of the identified NARXESN model on the validation dataset.}
    \label{fig:sim_NARXESN_validation}
\end{figure}

\vspace{-10pt}\[
\eta^* = \{ \Phi^{\textrm{sel},*},\, \nu = 6,\, n_{nl} = 3,\, \rho = 0.4535,\, k_z = 1.0178 \},
\]
with $$\Phi^{\textrm{sel},*} = \{ z(k),\, z(k-1),\, u(k),\, u(k-1)\},$$
and where the identified parameters are
\[
W_{out_1}^*= [\,-0.1011,\, 0.1585,\, -0.0702,\, 0.0251,\, -0.0699,\, -0.3192\,],\]
\[W_{out_2}^* = [\,0.1705,\, 0.2070,\ 0.3150,\, 0.2628\,].\]
As can be observed, unlike hyperparameter selection, parameter estimation is close but not perfect. Nevertheless, in the case of the NARXESN models, the choice of model structure appears to have a much stronger impact on the simulation performance than the accuracy of the estimated parameters.
In fact, the identified model demonstrated excellent performance on both training and validation datasets, as shown in Figure~\ref{fig:sim_NARXESN_validation}, achieving a FIT index of 95.39\% on the validation set.

\subsection{Discussion}

As illustrated in Section \ref{subsec:NARXESN_Results}, the proposed algorithm successfully identified the correct model structure, demonstrating robustness in the presence of measurement noise. While minor discrepancies were observed in parameter estimation, these did not significantly impact on the simulation performance. The absence of alternative state-of-the-art established structure selection algorithms for NARXESN models limits direct performance comparisons, but the achieved results are promising and indicate the potential of the proposed method in practical applications.

\section{Real-World Case Study Results}\label{sec:RealWorldCaseStudy}

In this section we evaluate the performance of the proposed structure selection algorithm on a real-world plant, specifically the Wiener–Hammerstein Benchmark system described in \cite{schoukens_wiener-hammerstein_nodate}. The objective is to assess the algorithm's ability to identify an accurate NARXESN model structure based on real experimental data.

\subsection{System Description}\label{sec:WHS_sys_description}

\begin{figure}[H]
    \centering
    \includegraphics[width=0.8\textwidth]{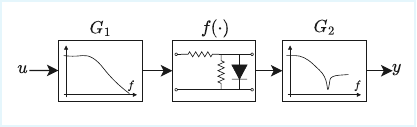}
    \caption{Block diagram of the Wiener–Hammerstein Benchmark system.}
    \label{fig:WHS_blockScheme}
\end{figure}

The Wiener–Hammerstein Benchmark is an electronic nonlinear system composed of three main elements: a linear dynamic block $G_1(s)$, a static nonlinear block $f(\cdot)$, and a second linear dynamic block $G_2(s)$, as depicted in Figure~\ref{fig:WHS_blockScheme}.

The system consists of:
\begin{itemize}
    \item $G_1(s)$: a third-order Chebyshev filter with a 0.5 dB pass-band ripple and a cut-off frequency of 4.4 kHz.
    \item $G_2(s)$: a third-order inverse Chebyshev filter with a 40 dB stop-band attenuation starting at 5 kHz.
    \item $f(\cdot)$: a static nonlinearity implemented using a diode circuit.
\end{itemize}

A band-limited Gaussian excitation signal (cut-off at 10 kHz) was applied, generating a dataset of 188,000 samples at a 51.2 kHz sampling rate.

\subsection{Model Structure Selection Results}\label{sec:WHS_structure-selection_results}

For model structure selection, a subset of $N=4000$ samples was used, equally divided into training and validation datasets.

The NARXESN structure selection algorithm was configured with an upper bound on the disturbance absolute value $\bar{w} = 5 \times 10^{-3}$. The candidate regressors were selected from the set:
\[
\Phi(10,10) = \{ z(k), z(k-1), \ldots, z(k-9), u(k), u(k-1), \ldots, u(k-9) \},
\]
resulting in $N_\Phi = 20$ regressors.

The algorithm considered the following ranges for the numerical hyperparameters:
\[
\nu \in [1, 15], \quad n_{nl} \in [0, \nu], \quad \rho \in (0,1), \quad k_z \in (0.5, 1.5).
\]

A total of $\mathcal{N}_s = 449$ scenarios were evaluated, with each scenario allowing a maximum of $\mathcal{I}_{MAX} = 20$ iterations. The minimum set-distance threshold was set to $J_{min} = 1 \times 10^{-6}$.

The final NARXESN model identified has the following hyperparameters:
\[
\nu^* = 11, \quad n_{nl}^* = 5, \quad \rho^* = 0.174, \quad k_z^* = 0.5252,
\]
and the regressors eventually selected are:
\begin{multline*}
\Phi^{\textrm{sel},*} = \{\, z(k),\, z(k-1),\, z(k-2),\, z(k-3),\, \\ u(k),\, u(k-2),\,
u(k-5),\, u(k-6),\, u(k-9) \,\}.
\end{multline*}
The estimated parameters are:
\begin{multline*}
W_{out_1}^* = [\, 0.0030,\, 0.1214,\, -0.0134,\, 0.0384,\, 0.0410,\, 0.0023,\, \\
0.0076,\, 0.0009,\, -0.0027,\, 0.0007,\, 0.0022 \,],
\end{multline*}
\begin{multline*}
W_{out_2}^* = [\, 2.2931,\, -1.2615,\, -0.3933,\, 0.5920,\, 0.0059,\, 0.0561,\, \\
0.0234,\, -0.0572,\, -0.0150 \,].
\end{multline*}

\begin{figure}[H]
    \centering
    \includegraphics[width=0.8\textwidth]{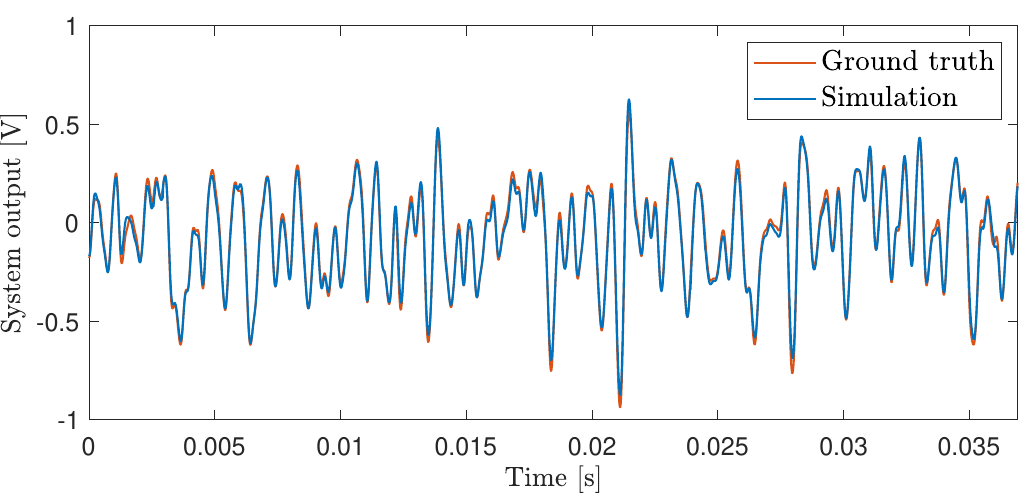}
    \caption{Simulation of the selected NARXESN model on the validation dataset.}
    \label{fig:WHS_NARXESN_validation}
\end{figure}

The model's performance on the validation dataset (see Figure~\ref{fig:WHS_NARXESN_validation}) achieves a fit index of \(\mathrm{FIT}\%=92.49\%\) and a root mean squared error \(\mathrm{RMSE}=18.1\,\mathrm{mV}\). 

\subsection{Performance Evaluation via Incremental Grid Search}\label{sec:WHS_grid_search}

To assess the effectiveness of the proposed structure–selection algorithm, we designed an alternative \textit{incremental grid–search procedure} to serve as a reference approach. 
Since, to the best of our knowledge, no comparable model–selection strategy specifically tailored for NARXESN models exists in the literature, this incremental procedure can be regarded as a reasonable empirical alternative that a practitioner could adopt for hyperparameter tuning.

The proposed comparison approach proceeds as follows. 
For a fixed subset of candidate regressors of \(u(\cdot)\) and \(z(\cdot)\), 
the best combination of numerical hyperparameters \(\nu^*, n_{nl}^*, \rho^*, k_z^*\) 
is determined by means of an exhaustive search over their admissible ranges. 
To prevent the combinatorial explosion resulting from the independent selection of each regressor, 
the autoregressive and exogenous input orders are constrained to be equal, i.e., \(n_u = n_z\), and all regressors up to lag \(n_z\) and \(n_u\) are included in the model, namely 
\[\Phi^\textrm{sel}=\Phi(n_u,n_z)=\{
z(k),\, \ldots,\, z(k-n_z+1),u(k),\, \ldots,\, u(k-n_u+1)\}.
\]

This simplification avoids the exponential growth that would otherwise arise from independently selecting each regressor, leading to \(2^{n_u+n_z}\) possible structures for each couple \((n_u,n_z)\).

The procedure then incrementally increases the model order according to
\[
n_u = n_z = 1,\, \ldots,\, 10,
\]
and, for each choice of \(n_u = n_z\), 
the remaining numerical hyperparameters are explored over the following ranges and resolutions:
\[
\begin{aligned}
\nu &\in [1,\,15], & \Delta\nu &= 2, \\
n_{nl} &\in [0,\,\nu], & \Delta n_{nl} &= 2, \\
\rho &\in (0,\,1), & \Delta\rho &= 0.2, \\
k_z &\in [0.5,\,1.5], & \Delta k_z &= 0.2.
\end{aligned}
\]
For each configuration, the model parameters were identified using the procedure described in Algorithm~\ref{alg:parametersIdentification}, 
and the corresponding validation performance was evaluated. 
Among all tested configurations within the specified ranges, 
the optimal combination \(\nu^*, n_{nl}^*, \rho^*, k_z^*\) was selected for each fixed pair \(n_u = n_z\) 
as the one achieving the minimum root mean squared error (RMSE) on the validation dataset.
\begin{table}[h!]
  \centering
  \caption{Incremental grid–search results for NARXESN models. 
  For each model order \(n_u = n_z\), the optimal hyperparameters 
  \(\nu^*, n_{nl}^*, \rho^*, k_z^*\) and the corresponding validation 
  performance (FIT and RMSE) are reported.}
  \label{tab:NARXESN_grid_search_results}

  \renewcommand{\arraystretch}{1.05}
  \setlength{\tabcolsep}{11pt}
  \rowcolors{2}{white}{gray!10}

  \begin{tabular}{c|cccccc}
    \toprule
    \(\bm{n_u=n_z}\) & \(\bm{\nu^*}\) & \(\bm{n_{nl}^*}\) & \(\bm{\rho^*}\) & \(\bm{k_z^*}\) & 
    \textbf{FIT (\%)} & \textbf{RMSE (mV)} \\
    \midrule
    1  & 13 & 1  & 0.05 & 0.8  & 87.39 & 30.99 \\
    2  & 15 & 1  & 0.85 & 1.2  & 87.67 & 30.30 \\
    3  & 9  & 1  & 0.25 & 1.0  & 87.87 & 29.82 \\
    4  & 11 & 4  & 0.05 & 0.8  & 87.94 & 29.65 \\
    5  & 5  & 4  & 0.85 & 1.4  & 87.97 & 29.58 \\
    6  & 3  & 1  & 0.05 & 0.6  & 88.03 & 29.42 \\
    7  & 13 & 4  & 0.45 & 1.2  & 88.28 & 28.80 \\
    8  & 15 & 4  & 0.25 & 0.6  & 88.11 & 29.23 \\
    9  & 11 & 10 & 0.25 & 0.6  & 87.81 & 29.95 \\
    10 & 15 & 10 & 0.25 & 1.4  & 88.64 & 27.93 \\
    \bottomrule
  \end{tabular}

  \rowcolors{0}{}{}
\end{table}
By incrementally increasing \(n_u = n_z = 1,\ldots,10\) 
and exploring the associated numerical hyperparameter space, 
a total of \(N_\mathcal{M} = 5750\) configurations were evaluated, 
corresponding to 575 models for each value of \(n_u = n_z\). 
The best-performing configuration for each order is summarized in Table~\ref{tab:NARXESN_grid_search_results}.
As reported in Table~\ref{tab:NARXESN_grid_search_results}, 
increasing the model order \(n_u = n_z\) generally leads to improved prediction accuracy, 
as reflected by higher FIT values and lower RMSE. 
The best configuration is obtained for \(n_u = n_z = 10\), 
yielding a validation performance of \(\mathrm{FIT} = 88.64\%\) and \(\mathrm{RMSE} = 27.93~\mathrm{mV}\). 
However, the overall performance gain tends to saturate for higher model orders, 
suggesting that additional regressors provide limited benefit once the main system dynamics are captured.

It is worth noting that the structures explored in this grid–search analysis 
are not directly comparable to those automatically selected by the proposed algorithm, 
as they rely on fixed regressor sets and equal input–output orders. 
Nevertheless, the results confirm that the automatically identified model 
achieves superior performance without requiring an exhaustive search. 
In addition, the incremental grid–search approach itself is not numerically tractable 
for higher–dimensional problems, 
since it rapidly suffers from the \textit{curse of dimensionality}, 
making exhaustive exploration of the hyperparameter space computationally infeasible.

\subsection{Training Method Comparison}\label{sec:training_comparison}

To evaluate the impact of the training procedure, i.e., Algorithm~\ref{alg:parametersIdentification}, on model accuracy, 
all model structures \(\mathcal{M}(\eta_i, \cdot)\), for \(i = 1, \ldots, \mathcal{N}_\mathcal{M}\), 
explored in the incremental grid–search (Section~\ref{sec:WHS_grid_search}) 
were identified using both the conventional Least Squares (LS) method 
and the proposed Algorithm~\ref{alg:parametersIdentification}. 
For each tested model, the resulting FIT and RMSE values 
were compared between the two estimation schemes.

\begin{figure}[h!]
    \centering
    \includegraphics[width=0.8\textwidth]{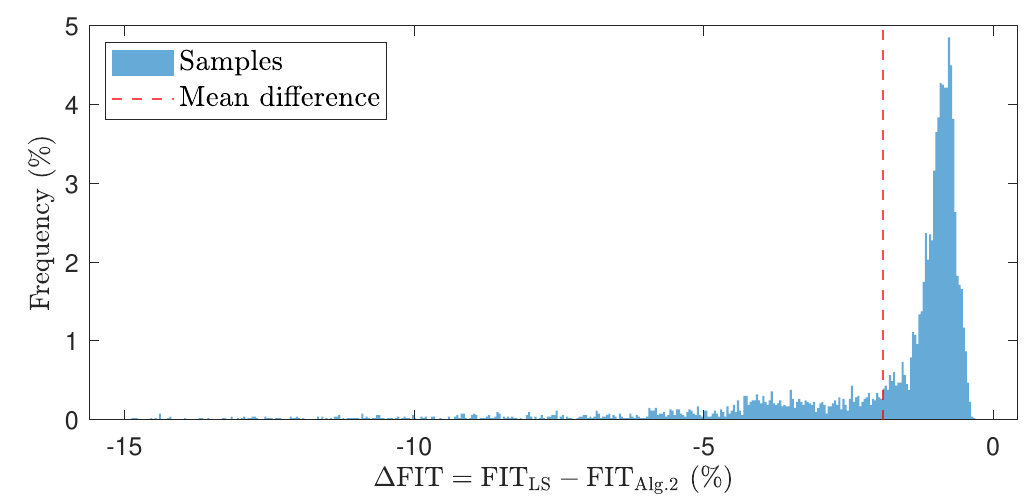}
    \caption{Histogram of FIT differences between LS and Algorithm~\ref{alg:parametersIdentification} over all grid–search configurations. 
    Negative values indicate higher FIT achieved by the proposed method.}
    \label{fig:WHS_hist_fit_diff}
\end{figure}

\begin{figure}[h!]
    \centering
    \includegraphics[width=0.8\textwidth]{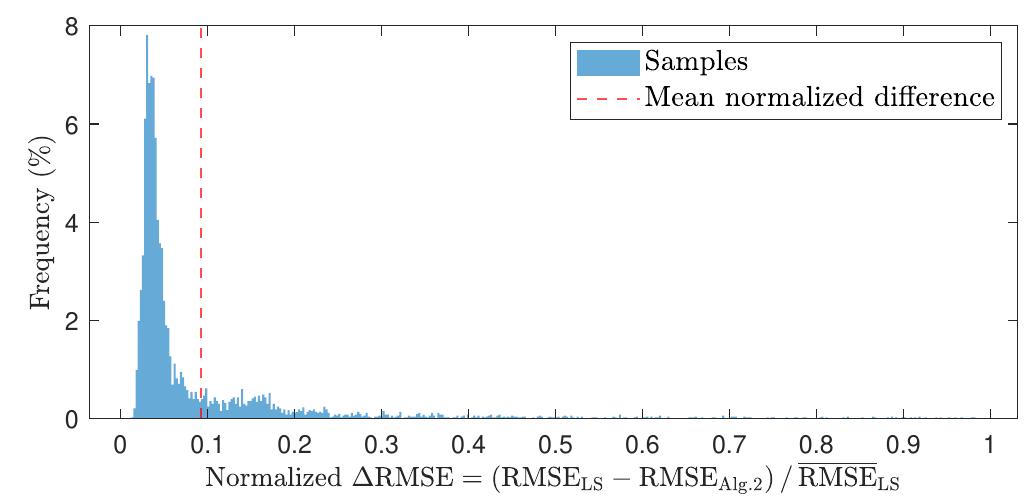}
    \caption{Histogram of normalized RMSE differences between LS and Algorithm~\ref{alg:parametersIdentification}.
    Positive values indicate lower RMSE achieved by the proposed method.}
    \label{fig:WHS_hist_rmse_diff}
\end{figure}

Figure~\ref{fig:WHS_hist_fit_diff} and Figure~\ref{fig:WHS_hist_rmse_diff} 
show the histograms of the performance differences 
computed over all grid–search configurations. 
The FIT difference is defined as 
\[
\Delta \mathrm{FIT} = \mathrm{FIT}_{\mathrm{LS}} - \mathrm{FIT}_{\mathrm{Alg.\ref{alg:parametersIdentification}}},
\]
whereas the RMSE difference is expressed in normalized form as 
\[
\Delta \mathrm{RMSE} = 
\frac{\mathrm{RMSE}_{\mathrm{LS}} - \mathrm{RMSE}_{\mathrm{Alg.\ref{alg:parametersIdentification}}}}
{\overline{\mathrm{RMSE}}_{\mathrm{LS}}},
\]
that is, relative to the average LS error $\overline{\mathrm{RMSE}}_{\mathrm{LS}}$ over all configurations. 
In this representation, negative values of \(\Delta \mathrm{FIT}\) 
and positive values of normalized \(\Delta \mathrm{RMSE}\) 
indicate superior performance of the proposed training algorithm.
The histogram of FIT differences (Figure~\ref{fig:WHS_hist_fit_diff}) 
shows values concentrated in the negative region, 
confirming that the proposed method consistently yields higher prediction accuracy, 
whereas the histogram of normalized RMSE differences 
(Figure~\ref{fig:WHS_hist_rmse_diff}) 
exhibits only positive values, highlighting a uniform reduction 
in model error magnitude.

The negative distribution of \(\Delta \mathrm{FIT}\) values 
and the strictly positive normalized \(\Delta \mathrm{RMSE}\) values confirm that, 
for every inspected configuration, 
the proposed algorithm systematically achieves higher FIT and lower normalized RMSE 
compared to LS estimation. 
This demonstrates a uniform enhancement in model accuracy 
when using the proposed training approach. 

\subsection{Discussion}

The analyses conducted with experimental data confirm the effectiveness and robustness of the proposed identification framework. 
The incremental grid–search results decribed in Section~\ref{sec:WHS_grid_search} highlight that, although larger model orders tend to improve accuracy, 
the gain quickly saturates while the computational burden increases significantly. 
In contrast, the proposed structure–selection algorithm automatically identifies a compact and accurate model 
without requiring exhaustive exploration of the hyperparameter space, 
thus overcoming the practical limitations imposed by the curse of dimensionality.

Moreover, the comparison between the LS and Algorithm~\ref{alg:parametersIdentification} training procedures 
demonstrates that the proposed method consistently yields higher FIT and lower RMSE values 
for every tested configuration, corresponding to an average error reduction of about 10\%. 
This confirms the advantage of the set–membership–based formulation 
in enhancing noise robustness and generalization capability.

Overall, the proposed approach provides a computationally efficient and reliable alternative 
to conventional tuning strategies, delivering parsimonious nonlinear models 
with high prediction accuracy suitable for real–world system identification tasks.

\section{Conclusions} \label{sec:Conclusions}

In this work we proposed a novel method for joint model structure selection and parameter learning for nonlinear dynamical systems, with a focus on NARX Echo State Networks (NARXESNs). The approach is based on set-membership identification and introduces a new selection criterion, the set-distance, used to compare candidate models and guide the iterative structure refinement process.

The method combines a robust parameter estimation step, relying on the computation of the feasible parameter set, with a scenario-based evaluation of model accuracy. Regressor selection is carried out by measuring the impact of each term on the set-distance, while redundant or non-informative features are pruned using the same metric. Numerical hyperparameters are tuned iteratively to further improve model fit.

The proposed algorithm was validated on both simulated and real-world systems. In the simulation case, the method correctly recovered the structure of the generating model. In the real-world benchmark, it achieved high prediction accuracy with a compact model, demonstrating its ability to handle noisy data and identify meaningful dynamics.

Future work may explore the application of this method to broader model families and to the identification of the structure and parameters of nonlinear controllers. The overall approach provides a flexible and reliable framework for nonlinear system identification with structure selection.

\section*{Acknowledgments}
\noindent The research activity of Corrado Sgadari has been supported by the Italian Ministry of University and Research (MIUR) under the PNRR program and by Ricerca sul Sistema Energetico S.p.A as co-financer of the grant.

\bibliographystyle{unsrt}
\bibliography{biblio} 

\end{document}